\documentclass[british]{article}

\usepackage[margin=0.7in]{geometry}
\usepackage[parfill]{parskip}
\usepackage[utf8]{inputenc}
\usepackage{amsmath,amssymb,amsfonts,amsthm}
\usepackage{amstext}
\usepackage{units}
\usepackage{float}
\usepackage{graphicx}
\usepackage[super,numbers]{natbib}

\usepackage{xr}
\usepackage[hidelinks]{hyperref}

\newcommand{\beginsupplement}{
        \setcounter{table}{0}
        \renewcommand{\thetable}{S\arabic{table}}%
        \setcounter{figure}{0}
        \renewcommand{\thefigure}{S\arabic{figure}}%
     }

\begin{document}

\title{Cascading parallel fractures on Enceladus}
\author{Douglas J. Hemingway{*}\textsuperscript{1,2}, Maxwell L. Rudolph\textsuperscript{3},
Michael Manga\textsuperscript{2}}

\maketitle
\textsuperscript{1}Department of Terrestrial Magnetism, Carnegie
Institution for Science, Washington, DC, 20015, \textsuperscript{2}Department
of Earth \& Planetary Science, University of California Berkeley,
Berkeley, CA, 94720, \textsuperscript{3}Department of Earth \& Planetary
Sciences, University of California Davis, Davis, CA, 95616. {*}Corresponding
author: dhemingway@carnegiescience.edu.

\section*{Main Text}

\textbf{Active eruptions from the south polar region of Saturn's small
($\sim\unit[500]{km}$ diameter) moon Enceladus are concentrated along
a series of lineaments known as the `tiger stripes' \citep{Porco2006,Porco2014},
thought to be partially open fissures that connect to the liquid water
ocean beneath the ice shell \citep{Ingersoll2016a,Kite2016}. Whereas
aspects of the tiger stripes have been addressed in previous work,
no study to date simultaneously explains why they should be located
only at the south pole, why there are multiple approximately parallel
and regularly spaced fractures, and what accounts for their spacing
of $\sim\unit[35]{km}$. Here we propose that secular cooling and
the resulting ice shell thickening and global tensile stresses \citep{Manga2007,Rudolph2009}
cause the first fracture to form at one of the poles, where the ice
shell is thinnest due to tidal heating \citep{Hemingway2019}. The
tensile stresses are thereby partially relieved, preventing a similar
failure at the opposite pole. We propose that subsequent activity
then concentrates in the vicinity of the first fracture as the steadily
erupted water ice loads the flanks of the open fissure, causing bending
in the surrounding elastic plate and further tensile failure in bands
parallel to the first fracture, leading to a cascading sequence of
parallel fissures until the conditions no longer permit through-going
fractures.}

The large amplitude of Enceladus' forced physical librations \citep{Thomas2016}
demonstrates that a global liquid water ocean is present beneath the
outer ice shell, consistent with the floating ice shell model proposed
to explain the muted gravitational field asymmetries \citep{Iess2014}
in spite of the large topography \citep{Collins2007,Nimmo2011a,Tajeddine2017}.
Interior models suggest a south polar shell thickness in the vicinity
of $\sim\unit[9]{km}$, whereas the north polar shell thickness is
likely closer to $\sim\unit[15]{km}$ \citep{Hemingway2019}. Given
that the erupted ice grains appear to be sampling this internal ocean
\citep{Postberg2009,Postberg2011}, the tiger stripes may be taken
as open fissures that fully penetrate the ice shell, providing a direct
conduit to the subsurface ocean. Accordingly, most studies presume
an extensional origin, although some features have been interpreted
as evidence of compressional or strike-slip activity \citep{Bland2015,Yin2016}.
What is least clear is how these fractures formed initially, why only
at the south pole, why as a subparallel set with regular spacing,
and why similarly active fissures have not yet been observed on other
icy bodies.

\begin{figure}
\begin{centering}
\includegraphics{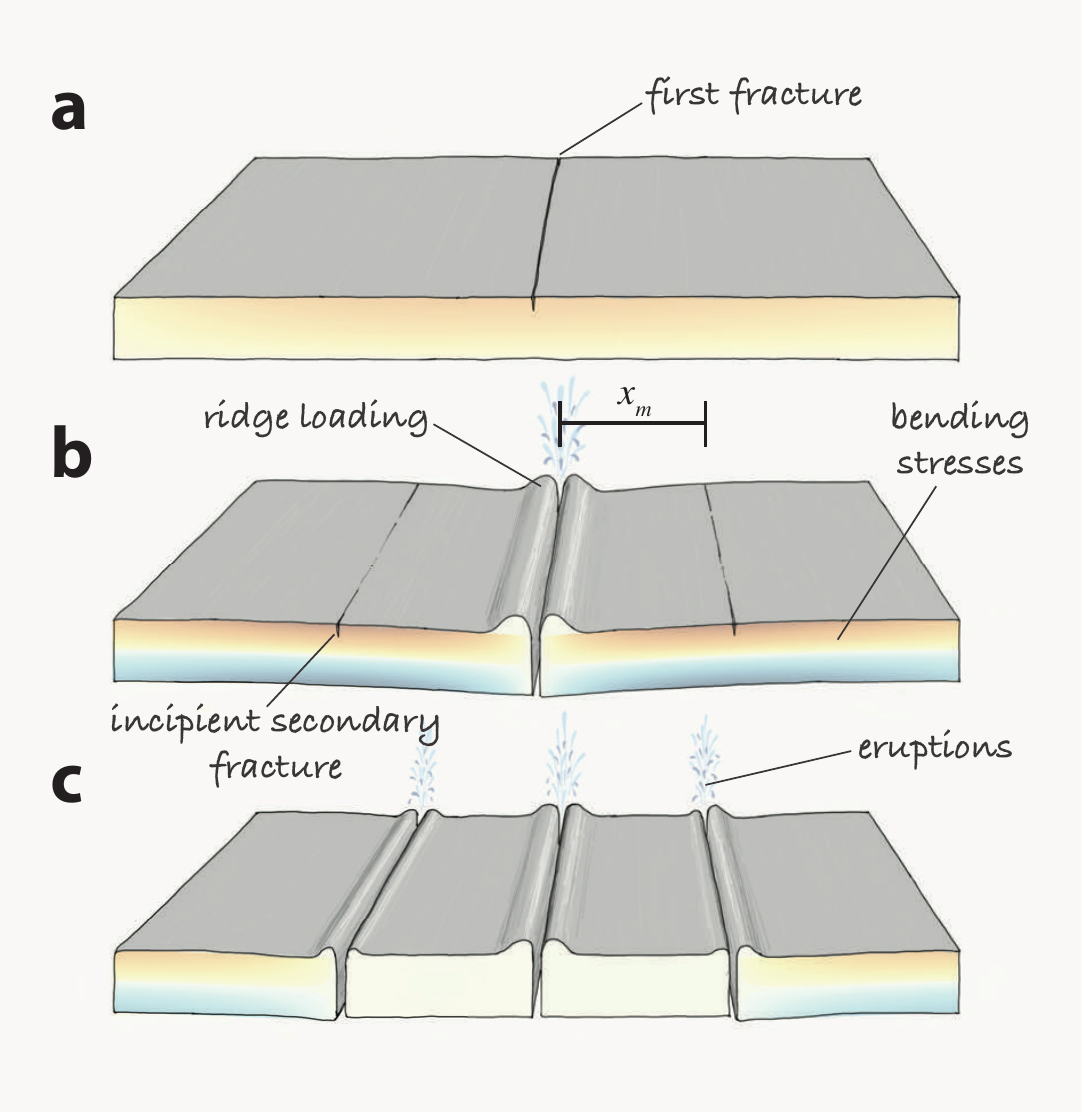}
\par\end{centering}
\caption{\label{fig:Cartoon-sequence}Schematic illustration of the proposed
sequence for tiger stripe formation. (a)~The global tensile stress
field (orange shading) arising from secular cooling leads to tensile
failure at one of the poles, where the ice shell is thinnest. (b)~Following
the first fracture's formation, the erupting solids then accumulate
preferentially in ridges flanking the open fissure, loading the edge
of the broken ice shell and producing bending stresses in the surrounding
elastic plate (respectively, orange and blue shades indicate tensile
and compressive stresses in the elastic layer; the ductile layer is
not shown). The bending stresses eventually become large enough to
initiate a set of secondary fractures parallel to the first and at
a distance, $x_{m}$, determined by the ice shell's elastic properties,
according to equation~(\ref{eq:xm_of_alpha}). (c)~Once open, the
secondary fractures then develop in a fashion similar to the first,
resulting in a cascading sequence of parallel fissures.}
\end{figure}

Secular cooling will result in some net freezing of the internal ocean.
As long as the overlying ice shell is sufficiently intact to support
tensile stresses on a global scale, the volume increase associated
with the phase change from water to ice will result in ocean pressurization
(\citealp{Manga2007}; Methods). As freezing progresses, tangential
stresses build until the tensile failure limit is exceeded somewhere
in the ice shell (Figure~\ref{fig:Cartoon-sequence}; Figure~\ref{fig:Secular-cooling-stresses}).
Since tidal heating should cause the ice shell to be thinnest at the
poles \citep{Choblet2017,Hemingway2019}, tensile stresses are maximized
at the poles such that the initial failure should occur at one of
the poles, with either pole being equally likely. Whereas the cold
upper part of the ice shell behaves elastically, the warmer ice toward
the base of the shell behaves viscously on long timescales (Methods).
Provided the ductile region of the ice shell is not too thick, however,
the fracture can rapidly penetrate the entire ice shell, establishing
an open pathway directly to the underlying ocean (\citealp{Rudolph2009};
Figure~\ref{fig:Initial-fracture-propagation}). Crucially, once
the first fissure forms, and provided that it remains open, the ice
shell is no longer capable of supporting global scale tangential stresses
and maintaining an over-pressurized ocean in this way. That is, it
is no longer possible for a similar fracture to develop at the opposite
pole, or anywhere else. We therefore suggest that Baghdad Sulcus,
which cuts directly through the geographic south pole, was the first
fracture to form and that the remaining fractures formed through a
distinct, though related, process (see below). Baghdad's orientation
of $\sim30{^\circ}$ from the tidal axis approximately maximizes normal
tensile stresses arising due to diurnal tidal deformation \citep{Nimmo2014,Nimmo2007}.
Although these tidal stresses are relatively weaker ($\sim\unit[14-85]{kPa}$),
when combined with the isotropic background tensile stress field resulting
from secular cooling, the total could be sufficient to cause tensile
failure. This may explain the orientation of the tiger stripes, provided
the ice shell has not experienced non-synchronous rotation since their
formation.

Once the first open fissure has formed, liquid water rises to the
level of neutral buoyancy (i.e., the lower $\sim90\%$ becomes filled
with water) and begins to boil off at the top, where it is exposed
to the vacuum. The rate of boiling may be controlled by back pressure
resulting from interaction between the vapour and the conduit walls
\citep{Nakajima2016} and turbulent dissipation associated with tidally
driven cyclic flushing and refilling of the liquid filled fissure
can prevent it from freezing shut \citep{Kite2016}. Provided these
(or similar) mechanisms can maintain the stability of such ocean-to-surface
pathways, the eruptions may persist for extended periods. Some 91\%
of the erupted solids fall ballistically back to the surface of Enceladus
\citep{Ingersoll2011}, accumulating preferentially on the flanks
of the open fissures \citep{Degruyter2011}. This effect may explain
the origin of the large ridges observed along the tiger stripes (Figure~\ref{fig:Cartoon-sequence};
\citealp{Dombard2013,Crow-Willard2015}). 

If the cold upper part of the ice shell behaves elastically, both
the material accumulated on the flanks of the open fissure and the
loss of buoyancy associated with any localised ice shell thinning
act like downward forces applied close to the edge of the broken elastic
plate. Such forces cause bending stresses to develop in the surrounding
ice shell (e.g., \citealp{TurcotteSchubert1982,Billings2005,Dombard2013};
Figure~\ref{fig:Elastic-plate-deflection}). Given a load acting
at the edge of the broken plate, it can be shown that the maximum
bending stresses occur at a distance from the first fracture given
by 
\begin{equation}
x_{m}=\alpha\frac{\pi}{4}\label{eq:xm_of_alpha}
\end{equation}
 where $\alpha$ is the characteristic length scale for elastic flexure,
given by 
\begin{equation}
\alpha^{4}=\frac{ET_{e}^{3}}{3\rho_{w}g\left(1-\nu^{2}\right)}\label{eq:alpha-explicit}
\end{equation}
where $E$ and $\nu$ are the Young's modulus and Poisson's ratio
for the icy shell, $g$ is the acceleration due to gravity, $\rho_{w}$
is the density of the ocean, and $T_{e}$ is the effective elastic
layer thickness. If the typical tiger stripe spacing of $\unit[\sim35]{km}$
is governed by the position where maximum bending stresses occur,
then, assuming typical values of $E=\unit[9]{GPa}$, $\nu=0.25$,
and $\rho_{w}=\unit[1000]{kg/m^{3}}$, and with $g=\unit[0.113]{m/s^{2}}$,
we obtain an elastic thickness of $T_{e}\approx\unit[5.2]{km}$ (Figure~\ref{fig:Maximum-tensile-stresses}).
Given the expected temperature structure of the ice shell, such an
elastic thickness corresponds to a total local shell thickness of
$\unit[\sim9]{km}$ (Methods), consistent with the value inferred
from gravity, topography, and librations \citep{Hemingway2019}. Approximating
the load as a line load, $V_{0}$, acting at the edge of the broken
elastic plate, it can be shown (Methods) that the resulting bending
causes maximum tensile stresses at $x_{m}$ to reach
\begin{equation}
\sigma_{\text{max}}=V_{0}\frac{6}{T_{e}^{2}}\alpha e^{-\pi/4}\sin\frac{\pi}{4}.\label{eq:sigma_max_main_text}
\end{equation}

Bending stresses can thus initiate secondary fractures in bands parallel
to the first fracture once the load acting at the edge of the plate
is sufficient to cause $\sigma_{\text{max}}$ to exceed the tensile
failure limit for ice, $\sigma_{\text{crit}}$. Although localised
ice shell thinning could, in principle, contribute to the bending
of the plate, the process may be self limiting because the locally
melting ice is opposed by viscous closure and the freezing that results
from reduced local turbulent dissipation, making this effect too small
to produce the necessary loading (Methods; Figure~\ref{fig:Melt-back-wedge};
\citealp{Kite2016}). A more effective source of edge loading may
be the accumulation of erupted material onto the ridges flanking the
open fissure. The eruption rate and the fraction of erupted material
falling back onto the ridges determines the time required before the
bending stresses cause tensile failure and the initiation of a secondary
fracture (Figure~\ref{fig:Fiber-stress-vs-time-and-mdot}). If, for
example, the eruption rate of the solids from a single fissure is
$\unit[20]{kg/s}$ and if 91\% of this material goes into the formation
of the flanking ridges, then, taking the fissure length to be $\unit[100]{km}$,
the loading on each ridge reaches the critical value after $\sim1$
million years assuming $\sigma_{\text{crit}}=\unit[1]{MPa}$, or $\unit[\sim100]{kyrs}$
if the ice fails at $\unit[100]{kPa}$ (Methods).

\begin{figure}
\begin{centering}
\includegraphics{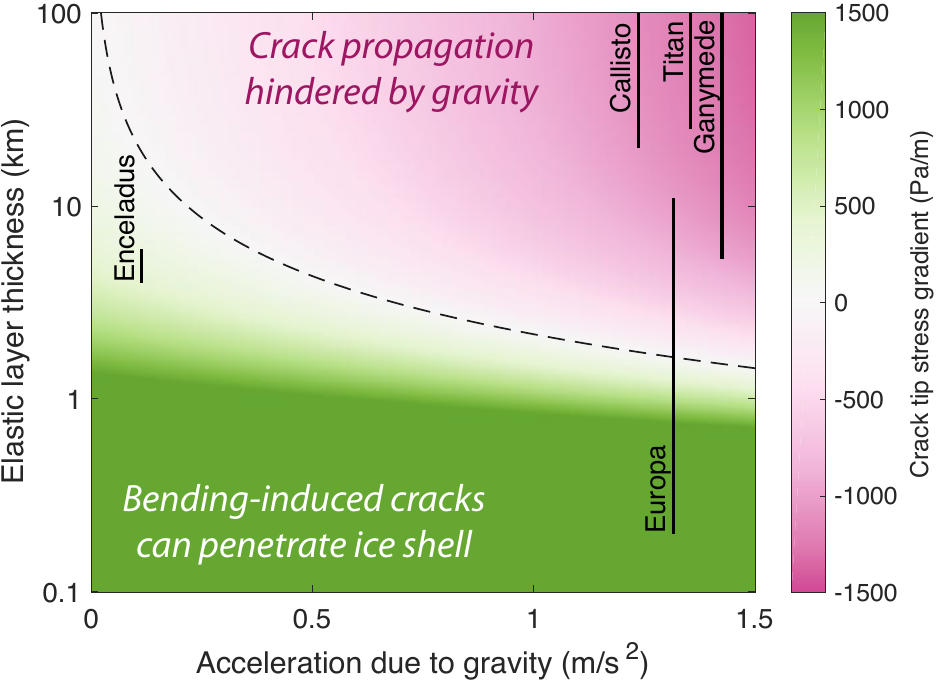}
\par\end{centering}
\caption{\label{fig:Crack-tip-stress-gradient}Crack tip stress gradient (change
in tip stress as the crack propagates) as a function of surface gravity
and effective elastic layer thickness, given by equation~(\ref{eq:sigma_tip_gradient}).
Positive values (green) correspond to conditions that permit the crack
tip stress to be increasingly tensile as the crack propagates. Negative
values (magenta) correspond to conditions in which the compressive
stresses due to overburden pressure build too rapidly to permit crack
propagation immediately following the initial bending-stress-induced
failure (see Figure~\ref{fig:Crack-tip-stress}). The dashed black
contour represents the transition between the two regimes. The vertical
black lines represent the estimated ranges of elastic layer thicknesses
for icy ocean worlds \citep{Spohn2003,Billings2005,Hemingway2013,Vance2014}.}
\end{figure}

As the secondary fracture begins to propagate, the broken portion
of the elastic plate can no longer contribute to supporting the bending
moment and the effective elastic thickness is reduced. Assuming the
load, $V_{0}$, is not changing on the timescale of fracture propagation,
the stress profile in the elastic plate becomes increasingly steep,
with the crack tip stress being always tensile and increasing steadily
as the fracture propagates. This effect is countered by the compressive
stresses from overburden pressure, which increase with depth (Figure~\ref{fig:Crack-tip-stress}).
For Enceladus, however, the gravity is so weak that such compressive
stresses are not important and the crack tip stresses are always increasing
as the crack propagates, meaning that the crack should penetrate rapidly
through the entire elastic layer. For larger icy ocean worlds, the
gravity is generally too strong to allow the crack tip stress to build
in this way, potentially explaining why similarly active fissures
are not observed on other bodies (Figure~\ref{fig:Crack-tip-stress-gradient};
Figure~\ref{fig:Crack-tip-stress}; Methods). Although roughly the
lower 40\% of the ice shell is ductile on long timescales (Methods),
it may behave elastically on the timescale of fracture propagation,
such that the secondary fracture can penetrate all the way through
to the ocean---though future work is required to model the full viscoelastic
nature of this problem. Once a through-going fracture is established
in this way, it evolves in a fashion similar to the first fracture,
allowing the sequence to cascade outward from the original fracture
in symmetric pairs (Figure~\ref{fig:Cartoon-sequence}): after Baghdad,
Cairo and Damascus, then Alexandria and the feature informally named
``E'' by \citet{Yin2015}.

There are several ways in which the cascade may be arrested. If the
eruption rate becomes too slow, bending stresses may not reach the
critical value on a timescale for which the ice behaves elastically.
If the eruption rate through each fissure decreases as new fissures
are formed, those forming later may not be loaded quickly enough for
the resulting bending stresses to generate additional fractures. Fractures
forming farther from the pole, where the background ice shell thickness
is greater, may also have a more difficult time propagating through
the thicker ductile part of the ice shell.

Finally, although our model may be able to account for the initial
formation of the tiger stripes, a number of other processes, such
as tidal and gravitational stresses, may continue to rework the fissures
forming strike-slip features \citep{Yin2015,Yin2016} or the funiscular
plains \citep{Bland2015}, for example. Likewise, our results are
not incompatible with stratigraphically older features \citep{Patthoff2011}
having formed through similar or distinct processes that may have
operated prior to the formation of the currently active fissures.

\subsection*{Data availability}

All required data are available in the published literature as indicated.

\subsection*{Code availability}

The computer code required to carry out the calculations discussed
herein is available upon request from the corresponding author.

\subsection*{Acknowledgements}

This work was made possible by the NASA/ESA Cassini mission to Saturn
and, in particular, the work of the Imaging Science Subsystem team.
We thank Robert Citron, Jacob Jordan, Simon Kattenhorn, Edwin Kite,
and Tushar Mittal for helpful discussions, and the CIDER working group
for early discussions that contributed to parts of this work. D.J.H.
was funded in part by the Miller Institute for Basic Research in Science
at the University of California Berkeley and in part by the Carnegie
Institution for Science in Washington, DC.

\subsection*{Author contributions}

M.M. developed the analytical equations for the secular cooling-induced
tangential stresses in the ice shell. M.R. developed the analytical
and boundary element models for crack penetration for the first fracture
and carried out the related calculations. M.R. computed the relationship
between turbulent dissipation in the fissure and the crack opening
angle. D.J.H. proposed the mechanism of forming subsequent parallel
fractures due to bending stresses and carried out the related calculations.
D.J.H. drafted the manuscript with inputs from M.M. and M.R. All authors
reviewed and commented on the manuscript.

\subsection*{Competing interests}

The authors declare no competing interests.

\beginsupplement

\section*{Methods\label{sec:Methods}}

\subsection*{Temperature structure}

The total ice shell thickness ($d$) and the effective elastic thickness
($T_{e}$) are related by the temperature structure of the ice shell.
For a conductive ice shell, accounting for the temperature-dependent
thermal conductivity, which goes as $T^{-1}$, the temperature structure
is given by (e.g.,~\citealp{Nimmo2004}) 
\begin{equation}
T\left(z\right)=T_{s}\left(\frac{T_{b}}{T_{s}}\right)^{z/d}\label{eq:Tz}
\end{equation}
where $z$ is the depth below the surface and where $T_{s}$ and $T_{b}$
are the surface and basal temperatures, respectively. On long timescales,
the relatively warmer parts of the ice shell behave viscously while
the coldest parts of the ice remain elastic. If the ductile layer
corresponds to the ice that is warmer than $\unit[160]{K}$, then
from (\ref{eq:Tz}), and assuming $T_{s}=\unit[75]{K}$ and $T_{b}=\unit[273]{K}$,
we obtain $T_{e}/d\approx0.59$.

\subsection*{Tensile stress accumulation due to secular cooling}

Here, beginning with equations (2)-(4) of \citet{Manga2007}, we obtain
a set of closed form analytical expressions that capture the effects
of ocean overpressure and ice shell tensile stress accumulation resulting
from secular cooling and the corresponding ice shell thickening. Following
the notation of \citet{Manga2007}, we assume an ice shell of outer
radius $R$ and inner radius $r_{i}$ sitting above a global liquid
water ocean over a rocky core of radius $r_{c}$. Taking the ice and
water densities to be $\rho_{i}$ and $\rho_{w}$, respectively, it
can be shown that the excess pressure in the ocean resulting from freezing a thickness
of water $h$ is given by

\begin{equation}
P_{\text{ex}}=\frac{h\left(1-\frac{\rho_{i}}{\rho_{w}}\right)}{\frac{\beta\left(r_{i}^{3}-r_{c}^{3}\right)}{3r_{i}^{2}}+\frac{\xi}{E}\left(1-2\nu\frac{1+\frac{1}{2}\left(\frac{R}{\xi}\right)^{3}}{\left(\frac{R}{\xi}\right)^{3}-1}\right)}\label{eq:P_ex}
\end{equation}
where $\xi$ is the radius corresponding to the base of the elastic
layer ($\xi=R-T_{e}$, if the elastic layer thickness is $T_{e}$).
This excess ocean pressure drives tangential stresses in the ice shell
that, when evaluated at the outer surface $r=R$, are given by

\begin{equation}
\sigma_{t}=\frac{3}{2}\frac{P_{\text{ex}}}{\left(\frac{R}{\xi}\right)^{3}-1}.\label{eq:sigma_t}
\end{equation}

\begin{figure}[H]
\begin{centering}
\includegraphics[width=9cm]{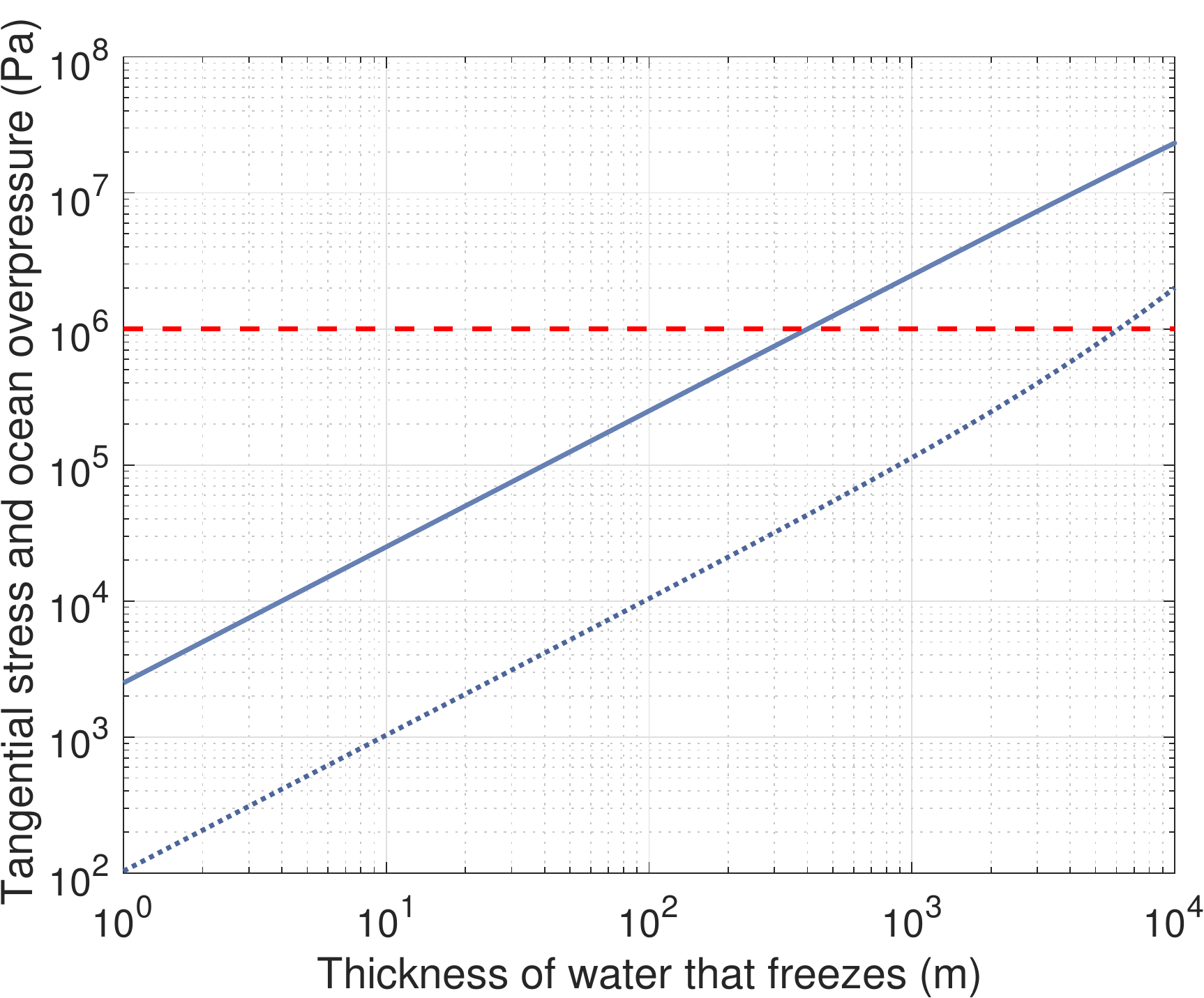}
\par\end{centering}
\caption{\label{fig:Secular-cooling-stresses}Accumulation of excess ocean
pressure (dotted blue line) and the resulting tangential stresses
in the ice shell (solid blue line) as a function of the thickness
of water that has frozen in the presence of an ice shell that is capable
of supporting global tensile stresses. Also shown is a nominal tensile
failure limit for cold ice (dashed red line).}
\end{figure}

Figure~\ref{fig:Secular-cooling-stresses} shows that, starting with
an intact ice shell that is capable of supporting global scale tangential
stresses, a few hundred meters of ocean freezing is sufficient to
cause tensile failure in the ice shell. This result is consistent
with the findings of \citet{Manga2007} but was obtained analytically
rather than numerically.

\citet{Manga2007} assumed a uniform ice shell such that there was
no preferred location for the tensile failure. For purposes of the
above calculation, we have made the same assumption. However, whereas
a non-uniform ice shell thickness should not significantly affect
the above result, it does affect the preferred location for the tensile
failure as stresses will concentrate in the thinnest parts of the
shell. Even prior to the initiation of significant eruptive activity,
the ice shell is expected to have been thinnest at the poles due to
the non-uniform distribution of tidal dissipation \citep{Hemingway2019}.
Hence, the tangential stresses resulting from secular cooling should
have been concentrated at both poles, with tensile failure being equally
likely to occur at either pole.

\subsection*{Propagation of initial fracture}

Once initiated at the surface (Figure~\ref{fig:Cartoon-sequence}a),
tensile cracks will propagate rapidly downward. We develop a model
based on linear elastic fracture mechanics (LEFM) to calculate the
maximum depth of penetration of fractures. We assume that on long
timescales, cold ice near Enceladus's surface supports elastic stresses
but warmer ice at the base of the ice shell behaves as a viscous fluid
and does not support elastic stresses (see above). As the subsurface
ocean is pressurised, tensile stresses are generated globally in the
elastic layer, and if these stresses exceed the tensile strength of
ice, a fracture will initiate at the surface and propagate downward.
The model from the previous section always predicts maximum tensile
stresses at the surface, rather than at intermediate depths within
the ice shell. If the tip of the fracture extends below the stressed
elastic layer, the upper region of the crack is under tension but
the tip of the crack is under compression due to overburden pressure.
We model the downward propagation of the fracture under the assumption
that the entire ice shell behaves as a linear elastic solid on the
timescale of crack propagation but that only the elastic layer supports
tensile stresses that encourage fracture penetration. We use a boundary
element code based on the displacement discontinuity method \citep{Crouch1983},
extended to include a crack-tip element \citep{Rudolph2009} to more
accurately resolve the stress field near the crack tip. In each calculation,
we initiate a short crack near the surface, extending to one-half
the depth ($d_{c}$) where lithostatic compression balances the applied
tensile stress ($\sigma_{t}$)

\begin{equation}
\int_{0}^{T_{e}}\sigma_{t}dz=\int_{0}^{d_{c}}\rho_{i}gzdz\label{eq:first_fracture_stress_balance}
\end{equation}
where $z$ is depth from the surface. We note that this formula is
similar to the maximum depth of fracture penetration from \citet{Qin2007},
except that we account for the presence of a viscous ice layer. We
solve for $d_{c}$, obtaining 
\begin{equation}
d_{c}=\sqrt{\frac{2T_{e}\sigma_{t}}{\rho_{i}g}}.\label{eq:dc}
\end{equation}

\begin{figure}[H]
\begin{centering}
\includegraphics[width=10cm]{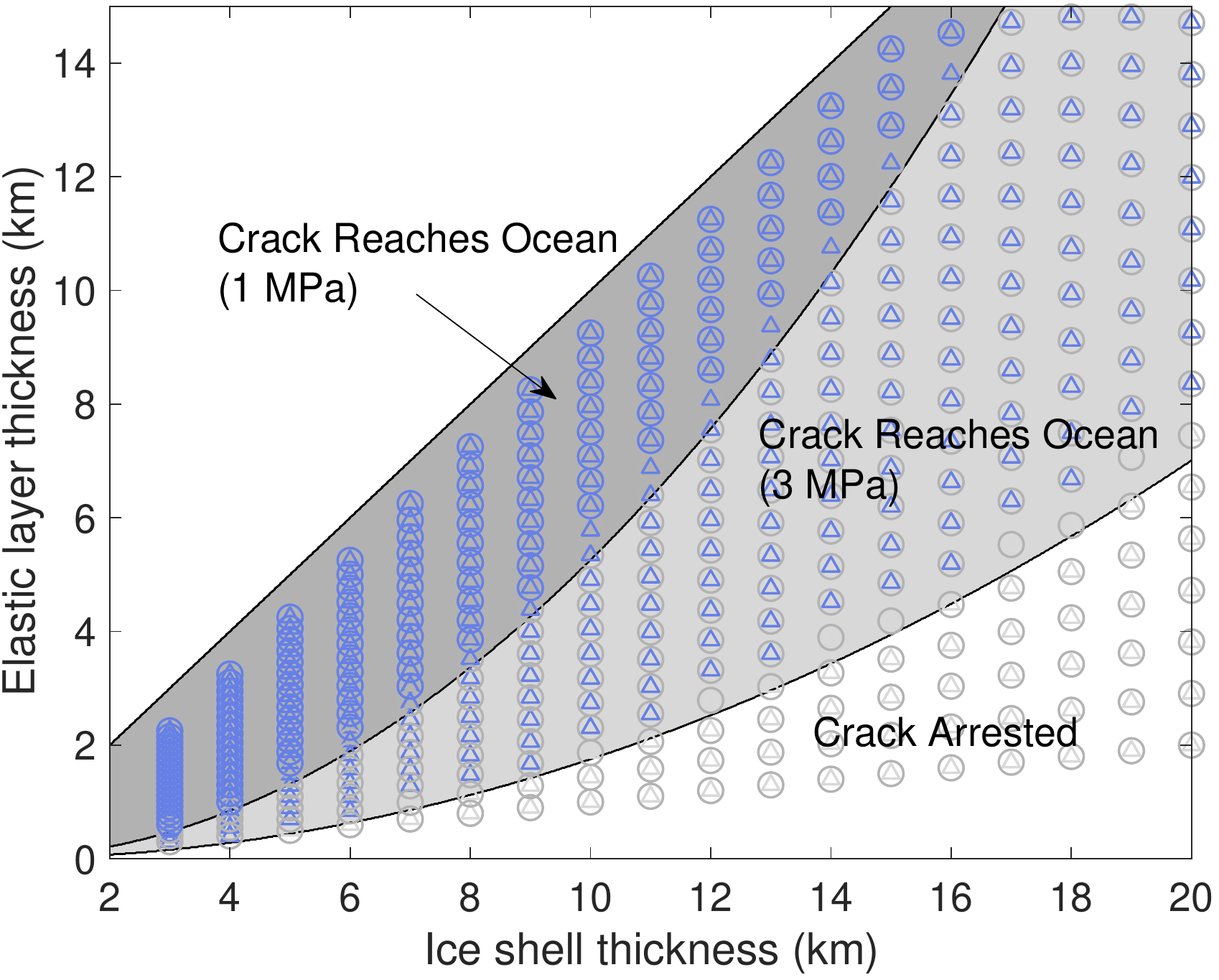}
\par\end{centering}
\caption{\label{fig:Initial-fracture-propagation}Conditions determining whether
or not a fracture initiated at the surface can penetrate entirely
through the ice shell, with the dark and light grey regions corresponding
to applied stresses of $\unit[1]{MPa}$ or $\unit[3]{MPa}$, respectively,
and delineated with the solid black lines given by equation~(\ref{eq:dc}),
assuming $g=\unit[0.113]{m/s^{2}}$ and $\rho_{i}=\unit[930]{kg/m^{3}}$.
The symbols represent the results of the numerical analysis, with
blue symbols indicating that the crack reaches the ocean and grey
symbols indicating that the crack is arrested; circles and triangles
correspond to applied stresses of $\unit[1]{MPa}$ or $\unit[3]{MPa}$,
respectively.}
\end{figure}

We calculate the mode-I stress intensity factor $K_{I}$ for the crack.
Next, we incrementally extend the crack length, computing a new solution
for displacements and $K_{I}$ for each successive crack length. The
crack is arrested if $K_{I}<0$, which is equivalent to assuming zero
fracture toughness. Our numerical and analytical results are in good
agreement with one another (Figure~\ref{fig:Initial-fracture-propagation})
and are consistent with the earlier work by \citet{Rudolph2009}.

\subsection*{Bending stresses}

The load acting at the edge of the broken plate (see main text and
below) causes bending stresses to develop in the elastic part of the
ice shell (e.g., \citealp{TurcotteSchubert1982,Billings2005}). We
assume the ice shell is completely broken at the initial fracture
and that there are no remaining horizontal loads---we assume that
membrane stresses are not important since they cannot be transmitted
across the open fissure. The elastic layer thickness is $T_{e}$ and
the total shell thickness is $d$. Approximating the problem in a
Cartesian geometry, and assuming cylindrical bending, the bending
moment at any given point is related to the curvature in the elastic
plate, and is given by
\begin{equation}
M\left(x\right)=D\frac{d^{2}w}{dx^{2}}\label{eq:M}
\end{equation}
where $w$ is the deflection, $x$ is the horizontal distance increasing
away from the fracture, and where $D$ is the flexural rigidity, given
by
\begin{equation}
D=\frac{ET_{e}^{3}}{12\left(1-\nu^{2}\right)}\label{eq:D}
\end{equation}
where $E$ and $\nu$ are the Young's modulus and Poisson's ratio
for the ice lithosphere. We assume the flexural rigidity does not
vary with $x$. The load is related to the deflection by
\begin{equation}
D\frac{d^{4}w}{dx^{4}}+\rho_{w}gw=q\left(x\right)\label{eq:load_equation}
\end{equation}
where $g$ is the acceleration due to gravity and $\rho_{w}$ is the
density of the water (we assume only vacuum exists above the ice shell).

For a line load acting at the edge of the broken plate, it can be
shown (e.g., \citealp{TurcotteSchubert1982}) that the deflection
is given by
\begin{equation}
w=\frac{V_{0}\alpha^{3}}{2D}e^{-x/\alpha}\cos\frac{x}{\alpha}\label{eq:w}
\end{equation}
where $V_{0}$ is the line load (in units of N/m) and where $\alpha$
is the characteristic length scale for flexure, given by
\begin{equation}
\alpha^{4}=\frac{4D}{\rho_{w}g}.\label{eq:alpha}
\end{equation}

It can be shown that the bending moment per unit length along the
fracture (units of N) is then given by
\begin{equation}
M\left(x\right)=V_{0}\alpha e^{-x/\alpha}\sin\frac{x}{\alpha}.\label{eq:M_of_alpha}
\end{equation}

The maximum bending moment occurs where $dM/dx=0$, at
\begin{equation}
x_{m}=\alpha\frac{\pi}{4}.\label{eq:xm}
\end{equation}

\begin{figure}
\begin{centering}
\includegraphics[width=9cm]{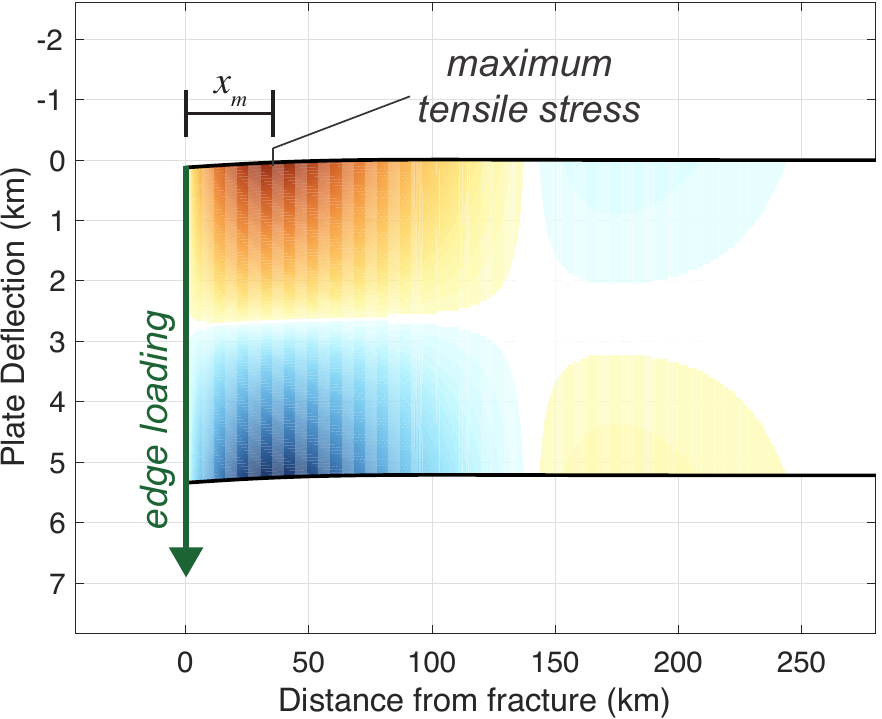}
\par\end{centering}
\caption{\label{fig:Elastic-plate-deflection}Elastic plate deflection and
internal stresses. Assumed parameters for this example are $E=\unit[9]{GPa}$,
$\nu=0.25$, $\rho_{w}=\unit[1000]{kg/m^{3}}$, and $g=\unit[0.113]{m/s^{2}}$,
for which the elastic thickness required to deliver $x_{m}=\unit[35]{km}$,
given by equations~(\ref{eq:xm_of_alpha}) and~(\ref{eq:alpha-explicit}),
is $T_{e}\approx\unit[5.22]{km}$. Deflection is computed according
to equation~(\ref{eq:w}) and is shown without any vertical exaggeration
(though note that the illustration's aspect ratio is approximately
30:1). Fibre stresses are computed with equation~(\ref{eq:sigma_xx})
and are illustrated with shading, where warm colours correspond to
tensile (positive) stresses and cool colours to compressive (negative)
stresses. Compressive stresses due to overburden pressure are not
shown. In the example illustrated, the deflection at the edge of the
plate reaches a maximum value of $\approx\unit[120]{m}$ when the
maximum fibre stress reaches $\unit[1]{MPa}$.}
\end{figure}

The fibre stresses (units of Pa) within the plate can be written
\begin{equation}
\sigma_{xx}\left(x,z\right)=\frac{12M\left(x\right)}{T_{e}^{3}}\left(\frac{T_{e}}{2}-z\right)\label{eq:sigma_xx}
\end{equation}
where $z$ is the vertical position measured downward from the top
of the plate (Figure~\ref{fig:Elastic-plate-deflection}). The maximum
fibre stresses occur at $x=x_{m}$, where the stresses are tensile
in the upper half (where $z<T_{e}/2$) and compressive in the lower
half (where $z>T_{e}/2$) of the deflected elastic plate.

\begin{figure}
\begin{centering}
\includegraphics[width=9cm]{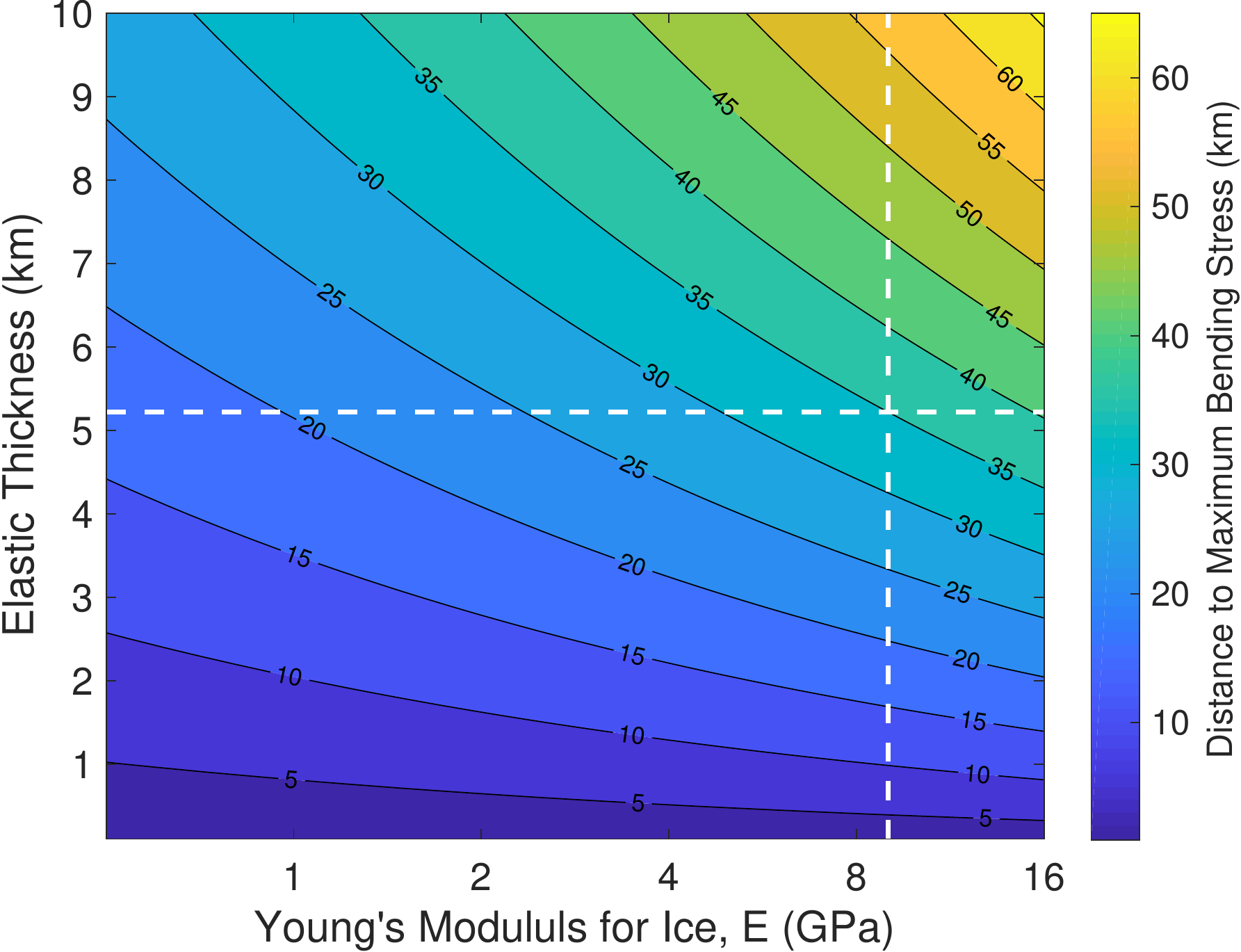}
\par\end{centering}
\caption{\label{fig:Maximum-tensile-stresses}Distance from the fracture to
the position of maximum tensile stress, computed via equation~(\ref{eq:xm}),
as a function of the ice shell's elastic properties: namely the Young's
modulus ($E$, shown on a logarithmic scale) and the effective elastic
layer thickness ($T_{e}$). A fixed Poisson's ratio of $\nu=0.25$
is assumed. As a point of reference, a nominal Young's modulus of
$E=\unit[9]{GPa}$ is indicated with a vertical dashed white line,
illustrating that the spacing of $\unit[35]{km}$ corresponds to an
elastic thickness of $T_{e}\approx\unit[5.2]{km}$, indicated by the
horizontal dashed white line.}
\end{figure}

This distance, $x_{m}$, varies as a function of the elastic properties
of the shell and is a 3/4 power function of the elastic thickness
($T_{e}$) and a 1/4 power function of the Young's modulus ($E$)
(Figure~\ref{fig:Maximum-tensile-stresses}). Assuming uniform properties
across the ice shell, tensile failure will occur at this distance
and parallel to the first fracture. If the typical tiger stripe spacing
of $\unit[\sim35]{km}$ is governed by the position of maximum bending
stresses, then this spacing can be used to determine the elastic properties
of the ice shell at the time of the fracture's formation. Given a
Young's modulus of $E=\unit[9]{GPa}$, for example, we find that the
maximum stresses occur at a distance of $\unit[35]{km}$ from the
fracture when the elastic thickness is $T_{e}\approx\unit[5.2]{km}$
(Figures~\ref{fig:Elastic-plate-deflection} and \ref{fig:Maximum-tensile-stresses}).
Assuming a predominantly conductive ice shell, and taking the elastic
layer to correspond to the uppermost part of the ice shell where the
temperature is $\unit[<160]{K}$ (see above), this elastic thickness
implies a total shell thickness (elastic plus ductile layers) of $\unit[\sim8.9]{km}$,
consistent with the polar ice shell thickness inferred from shape,
gravity, and libration observations \citep{Hemingway2019}. We note
that, whereas $\unit[\sim5.2]{km}$ may have been the relevant effective
elastic layer thickness at the time of the formation of the tiger
stripes, subsequent reworking of the ice shell may have introduced
faults that have reduced the modern effective elastic thickness to
perhaps $\unit[<2]{km}$, in line with estimates based on studies
of local flexural and tectonic features (e.g., \citealp{Bland2007,Giese2008}).

The elastic properties similarly determine how the magnitude of the
fibre stresses is related to the magnitude of the load. Given that
the maximum tensile stress occurs at $x=x_{m}$ and $z=0$, from equation
(\ref{eq:sigma_xx}), its magnitude is
\begin{equation}
\sigma_{\text{max}}=\frac{6M\left(x_{m}\right)}{T_{e}^{2}}\label{eq:sigma_max}
\end{equation}
which can also be written in terms of the required load, $V_{0}$,
as

\begin{equation}
\sigma_{\text{max}}=V_{0}\frac{6}{T_{e}^{2}}\alpha e^{-\pi/4}\sin\frac{\pi}{4}.\label{eq:sigma_max_of_V0}
\end{equation}

Tensile failure occurs when the load is sufficient to make the bending
moment at $x=x_{m}$ equal to the critical value of
\begin{equation}
M_{\text{crit}}=\frac{\sigma_{\text{crit}}T_{e}^{2}}{6}\label{eq:M_crit}
\end{equation}
where $\sigma_{\text{crit}}$ is the tensile failure limit for cold
ice, which we take to be $\unit[1]{MPa}$ (e.g.,~\citealp{Hammond2018})
for intact ice or $\unit[100]{kPa}$ for previously weakened ice.

\subsection*{Ice shell thinning}

\begin{figure}
\begin{centering}
\includegraphics[width=10cm]{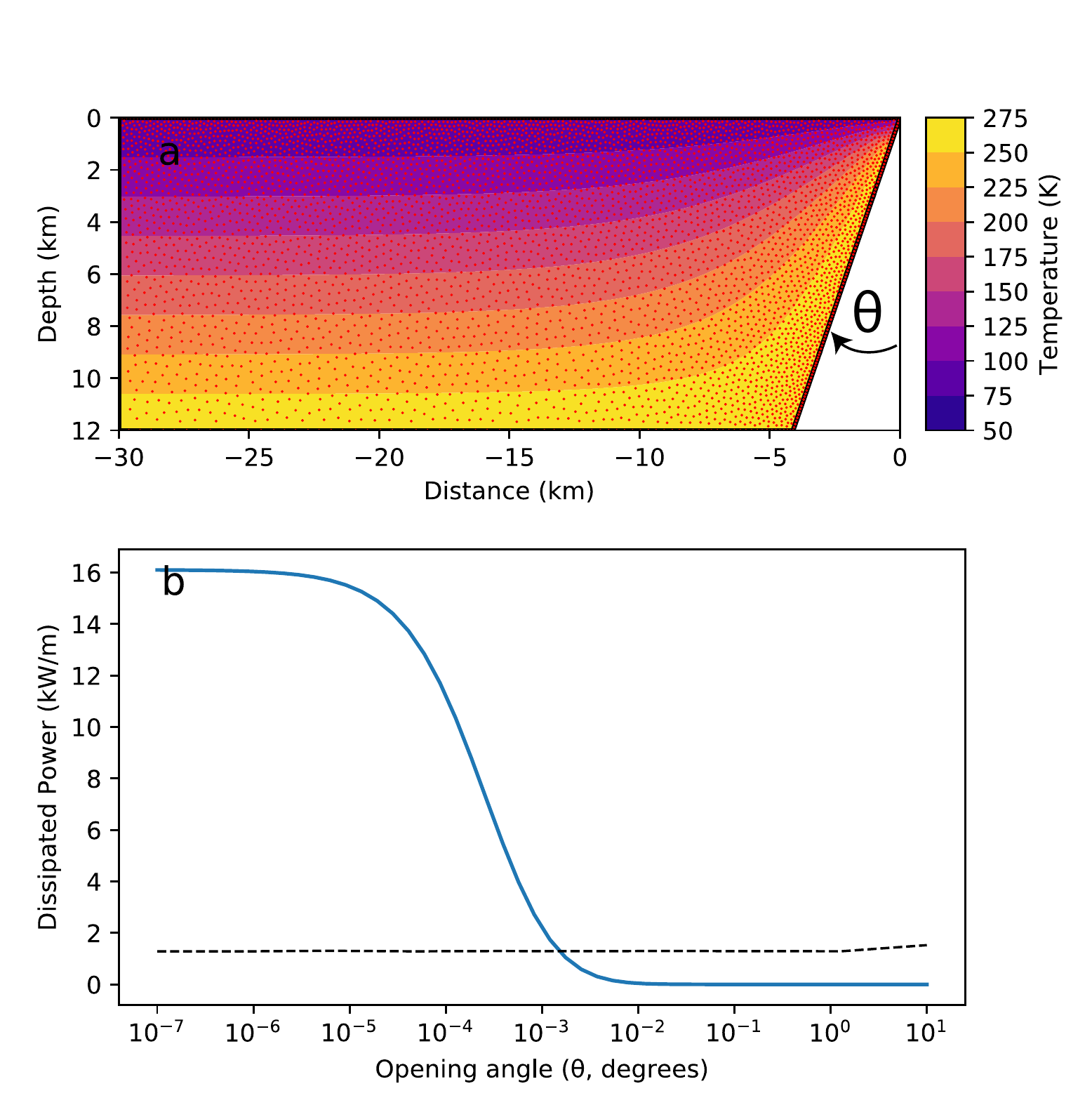}
\par\end{centering}
\caption{\label{fig:Melt-back-wedge}(a)~Development of ice shell temperature
structure in the vicinity of a liquid filled fissure. The red dots
represent the evaluation points in the radial basis function calculation.
(b)~Turbulent dissipation as a function of the fissure's opening
angle, $\theta$. The dashed black lines indicate the approximate
range of dissipation required to keep the fissure from freezing shut.
For opening angles larger than $\sim\unit[10^{-3}]{degrees}$, the
dissipation is not sufficient to keep the fissure open.}
\end{figure}

Here, we assess the possible effect of loss of buoyancy due to thinning
ice in the vicinity of an open fissure. We calculate the steady-state
temperature distribution around a fissure using a radial basis function
finite-difference approach. We use an isothermal surface boundary
condition of $T_{s}=\unit[75]{K}$, and an isothermal boundary condition
of $T_{m}=\unit[273]{K}$ on the crack wall and along the ice-ocean
interface (Figure~\ref{fig:Melt-back-wedge}a). To resolve the discontinuous
change in temperature where the crack meets the surface, we applied
a boundary condition along the crack that varies linearly from the
surface temperature to $\unit[273]{K}$ over the upper 10\% of the
ice shell thickness (i.e., the portion of the crack that extends above
the level of neutral buoyancy). We applied a far-field insulating
boundary condition at a distance $\unit[30]{km}$ from the crack.
For purposes of this calculation, the thermal conductivity was taken
as a constant ($\unit[2.5]{W/m/K}$). For a vertical crack in an ice
shell with thickness $\unit[12]{km}$, the steady-state conductive
heat flow is $\unit[\sim750]{W/m}$ per side. 

Once open, the tendency for the fissure to narrow due to freezing
is opposed by dissipation as water is cyclically flushed in and out
of the fissure (e.g., \citealp{Kite2016}). Since the temperature
gradient, and therefore the rate of conductive heat loss away from
the fissure, is greatest near the surface and decreases towards the
base of the fissure, the freezing will be most rapid at the top of
the fissure and slower toward its base. We idealise the crack as a
wedge shape and estimate the dissipation within the crack as well
as the heat conducted away from the crack.

We estimate the steady state opening angle, $\theta$, for the melt-back
wedge illustrated in Figure~\ref{fig:Melt-back-wedge}a. Assuming
the crack opens and closes periodically \citep{Hurford2007a}, we
can write the width of the crack, $b$, as a function of depth, $z$,
and time, $t$, as 
\begin{equation}
b\left(z,t\right)=b_{0}+2z\tan\theta+A\sin\left(\omega t\right)\label{eq:b_of_z_and_t}
\end{equation}
where $b_{0}$ is the width of the crack at the surface, $A$ is the
amplitude of the oscillations, and $\omega$ is the angular frequency.
Conservation of mass requires that the velocity of the liquid in the
crack, $u$, be related to the variations in crack width:
\begin{equation}
\frac{\partial\left(ub\right)}{\partial z}=\frac{\partial b}{\partial t}.\label{eq:conservation-of-mass}
\end{equation}

The energy dissipated by oscillatory flow in a fracture is found by
relating the resolved shear stress on the wall, $\tau$, to the mean
flow velocity, $<u>$, calculated from (\ref{eq:conservation-of-mass}),
using the Darcy-Weisbach equation
\begin{equation}
\tau=f\frac{1}{2}\rho<u>^{2}\label{eq:tau}
\end{equation}
where $f$ is a friction factor and $\rho$ is the fluid density.
While highly turbulent, the average properties of the flow are in
quasi-steady state, and the net acceleration of the water is negligible.
The wall shear stress is thus balanced by a dynamic pressure loss
per unit depth as 
\begin{equation}
2\tau=b\frac{dP}{dz}.\label{eq:tau_balance_dPdz}
\end{equation}

In turn, the rate of energy dissipation per unit volume ($Q_{v}$)
is related to the rate of decrease in dynamic pressure, 
\begin{equation}
Q_{v}=<u>\frac{dP}{dz}.\label{eq:dE_by_dV}
\end{equation}

We multiply $Q_{v}$ by the crack width to obtain the energy dissipated
per unit area of crack wall 
\begin{equation}
q_{d}=f\rho<u>^{3}.\label{eq:qd}
\end{equation}

Dissipation decreases very rapidly with increasing crack width (i.e.,
increasing opening angle). In Figure~\ref{fig:Melt-back-wedge}b,
we show predicted values of the dissipation integrated along the wall
of the crack (from the surface to the ocean-ice interface) for a crack
with a surface width of $\unit[1]{m}$ and an oscillation amplitude
of $\unit[0.9]{m}$, assuming a friction factor $f=0.01$ (representative
of fully-developed turbulent flow in a smooth channel). Owing to the
large uncertainties in the crack width, amplitude of oscillation,
and friction factor, the dissipation is also very uncertain, but our
result nevertheless demonstrates that if the crack begins to freeze
shut, dissipation in the crack becomes much larger than the conducted
heat flow, causing melting. On the other hand, dissipation alone cannot
produce enough melt-back to achieve opening angles larger than a few
hundredths of a degree. Thus, the dissipation mechanism, while capable
of preventing the fissure from freezing out (see also \citealp{Kite2016}),
does not produce a melt-back wedge wide enough to contribute significantly
to the bending stresses discussed above.

\subsection*{Ridge accumulation}

Erupted material can accumulate in ridges at the surface, loading
the edge of the broken elastic plate from the top. Such a load, per
unit length along the fracture, is 
\begin{equation}
V_{0}=\frac{mg}{L}\label{eq:V0}
\end{equation}
where $m$ is the mass of accumulated material and $L$ is the total
length of the ridges. Given an accumulation rate of $\dot{m}$, the
load as a function of time is given by
\begin{equation}
V_{0}\left(t\right)=\frac{\dot{m}g}{L}t.\label{eq:V0_of_t}
\end{equation}

\begin{figure}
\begin{centering}
\includegraphics[width=9cm]{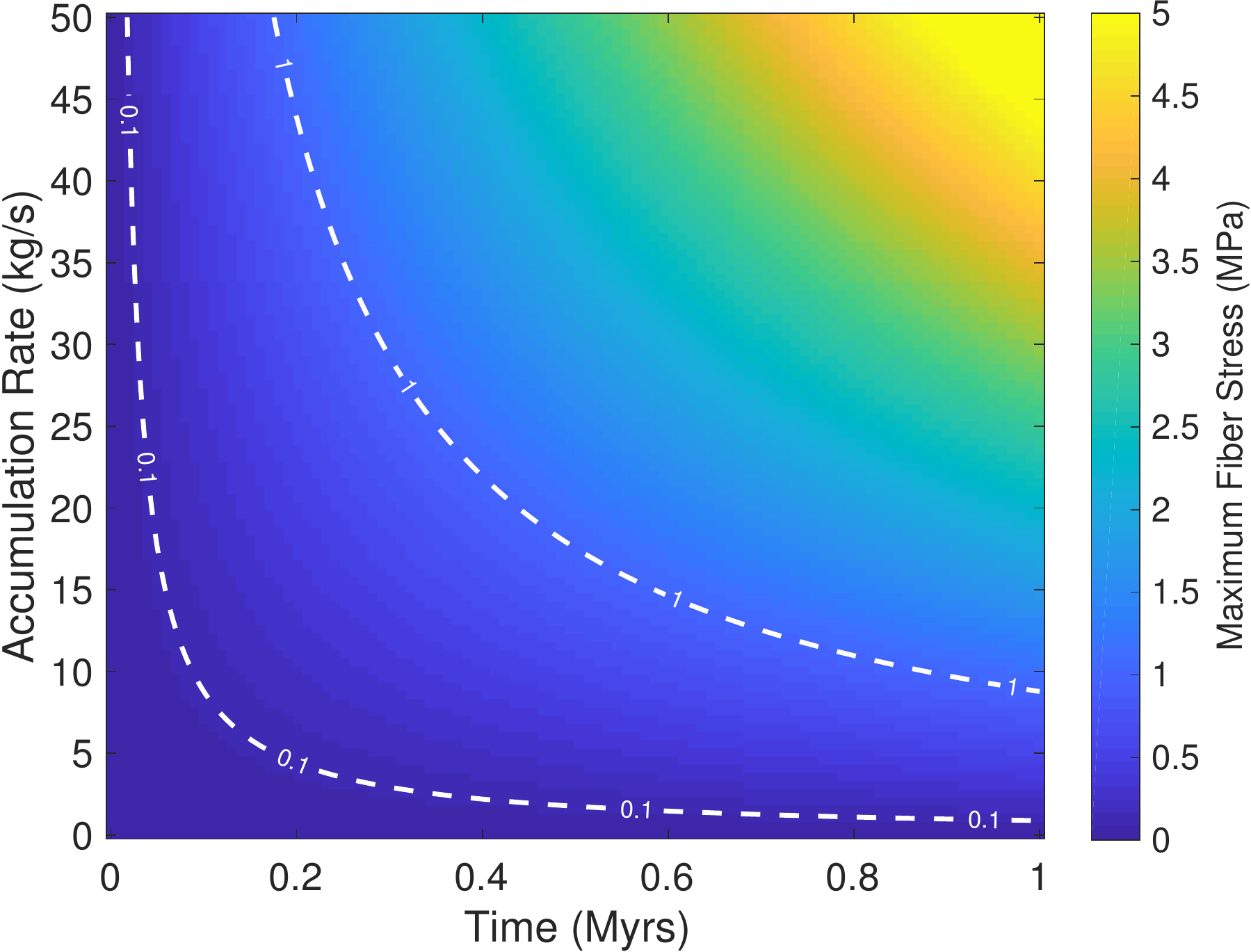}
\par\end{centering}
\caption{\label{fig:Fiber-stress-vs-time-and-mdot}Accumulated tensile stresses
at $x_{m}=\unit[35]{km}$ as a function of time and the rate of material
accumulation in the flanking ridge, whose length is assumed to be
$\unit[100]{km}$. The dashed white contours illustrate failure envelopes
assuming tensile failure limits of $\unit[100]{kPa}$ or $\unit[1]{MPa}$.}
\end{figure}

The accumulated load gives rise to tensile stresses which are maximum
at $x=x_{m}$ and given by equation~(\ref{eq:sigma_max_of_V0}).
Figure~\ref{fig:Fiber-stress-vs-time-and-mdot} shows this maximum
tensile stress as a function of time and accumulation rate, with contours
indicating two examples of tensile failure limits. Assuming a tensile
failure limit of $\unit[1]{MPa}$ and the same elastic properties
used in Figure~\ref{fig:Elastic-plate-deflection}, failure occurs
when $V_{0}=\unit[\sim3.1\times10^{8}]{N/m}$. Assuming an accumulation
rate of $\unit[10]{kg/s}$ per $\unit[100]{km}$ length of ridge,
the accumulated load will be sufficient to initiate tensile failure
after $\sim\unit[875]{kyrs}$. We note, as a point of reference, that
the current total rate of erupted solids has been estimated at $\unit[51\pm18]{kg/s}$
of which 9\% is estimated to escape from Enceladus while the remaining
91\% falls ballistically back to the surface \citep{Ingersoll2011}.
The rate of material accumulation per ridge is of course somewhat
smaller and depends on how one assumes the erupted material is distributed
among the tiger stripes, whose rates of activity vary. Note that the
eruption rates also vary with time over various timescales \citep{Nimmo2014,Ingersoll2017}.

The accumulated load may be related to the cross-sectional area, $A$,
of the ridge by
\begin{equation}
V_{0}=A\rho_{r}g\label{eq:V0_of_A}
\end{equation}
where $\rho_{r}$ is the assumed density of the ridge. Assuming $\rho_{r}=\unit[900]{kg/m^{3}}$,
for example, we obtain a load sufficient to generate tensile stresses
of $\unit[1]{MPa}$ when the cross-sectional area of the ridge is
$\unit[\sim3]{km^{2}}$. If the effective tensile failure limit at
the time of the tiger stripes' formation is only $\unit[100]{kPa}$,
then the required ridge cross section is just $\unit[\sim0.3]{km^{2}}$.
Given estimates of the height ($\unit[\sim150]{m}$) and width ($\unit[\sim1.5]{km}$)
of the present day ridges \citep{Crow-Willard2015} and accounting
for the deflection ($\unit[\sim120]{m}$) discussed above, the ridge
cross-sections may only be $\unit[\sim0.2]{km^{2}}$ at present. This
suggests that our proposed mechanism requires either that the bending-induced
failures occurred only after the effective tensile failure limit was
reduced to $\unit[\sim100]{kPa}$, perhaps due to pervasive weakening
following the formation of the first fissure, or that the ridges were
initially larger and have experienced some relaxation or erosion since
their formation.

\subsection*{Subsequent fractures}

\begin{figure}
\begin{centering}
\includegraphics[width=14cm]{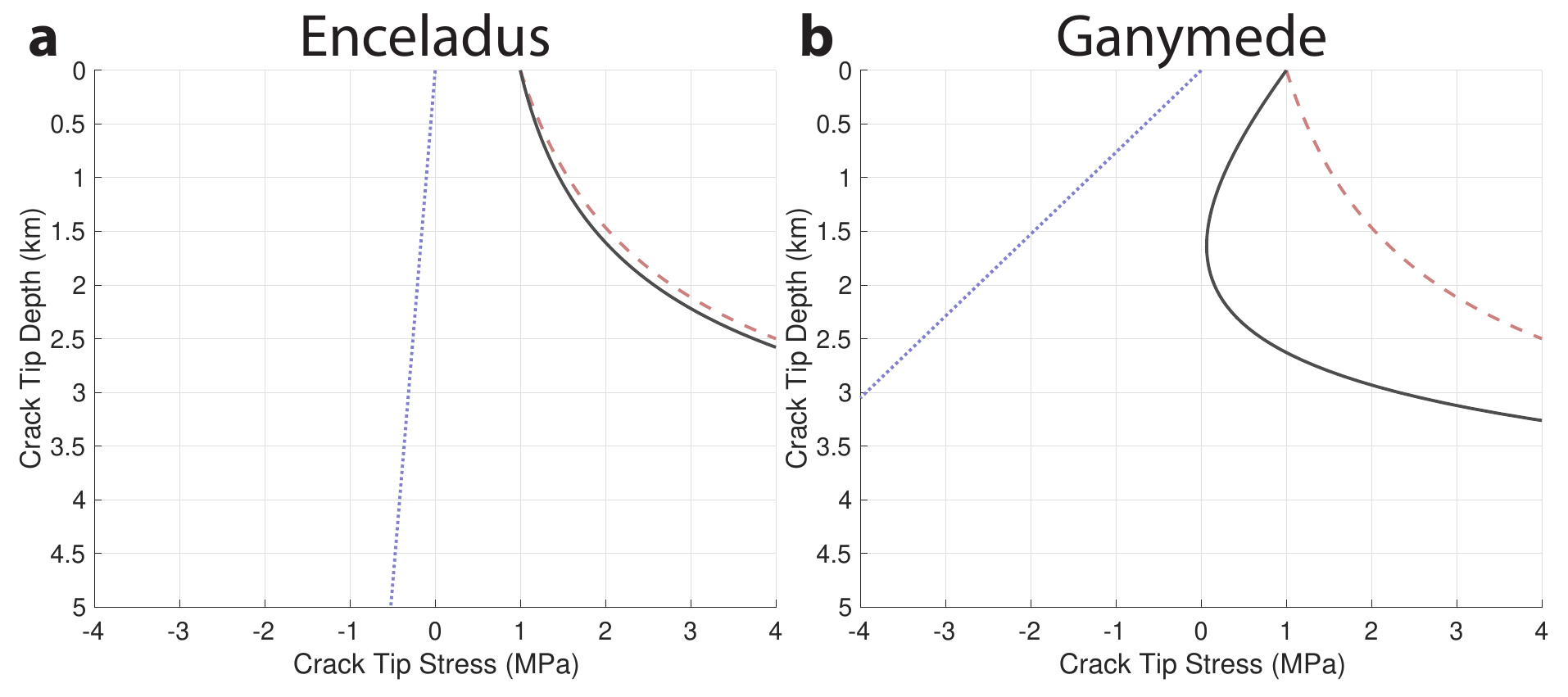}
\par\end{centering}
\caption{\label{fig:Crack-tip-stress}Stresses at the crack tip as a function
of crack tip depth for the cases of (a)~Enceladus and (b)~Ganymede.
Positive stresses are tensile, negative stresses are compressive.
The dotted blue line represents the compressive stresses due to lithostatic
pressure. The dashed red line represents the bending stresses given
by equation~(\ref{eq:sigma_tip_bending}). The solid black line represents
the net stress given by equation~(\ref{eq:sigma_tip}). Whereas for
Enceladus, the crack tip stress is always increasing as the crack
propagates downward, the crack tip stress initially decreases for
Ganymede, even when we assume an equally thin ice shell. The surface
gravity is $g=\unit[0.113]{m/s^{2}}$ for Enceladus and $g=\unit[1.428]{m/s^{2}}$
for Ganymede. For both examples, we assume $T_{e}=\unit[5.2]{km}$,
$\sigma_{\text{crit}}=\unit[1]{MPa}$, and $\rho_{\text{ice}}=\unit[920]{kg/m^{3}}$.}
\end{figure}

Once the bending stresses are great enough to cause tensile failure
at the surface, a new fracture is initiated (Figure~\ref{fig:Cartoon-sequence}b).
Whether or not the new fracture can penetrate the ice shell depends
on how the stress field evolves during crack propagation. Although
this is a dynamic process, we can gain some insight by considering
the hypothetical static scenario in which the crack is arrested after
propagating a distance $y$ from the surface. At this point, the load
($V_{0}$), which has not changed on the timescale of crack propagation,
would now have to be supported by the partially fractured plate. That
is, focusing on the location of the fracture, where $x=x_{m}$, the
bending moment that must be supported remains fixed at $M_{\text{crit}}$,
given by equation~(\ref{eq:M_crit}). The fibre stresses in the remaining
unbroken part of the lithosphere would have to support the same bending
moment but with a reduced effective elastic layer thickness. The bending
stress at the crack tip is thus
\begin{equation}
\sigma_{\text{tip,bending}}\left(y\right)=\frac{6M_{\text{crit}}}{\left(T_{e}-y\right)^{2}}=\frac{\sigma_{\text{crit}}T_{e}^{2}}{\left(T_{e}-y\right)^{2}}\label{eq:sigma_tip_bending}
\end{equation}
where $T_{e}$ is the initial elastic layer thickness (before the
fracture develops) such that the effective remaining elastic layer
thickness is always $T_{e}-y$. The bending stresses at the crack
tip are therefore always tensile and increasing in magnitude as the
crack tip propagates downward through the ice shell (dashed red lines
in Figure~\ref{fig:Crack-tip-stress}).

This effect is opposed by the background compressive stresses that
exist due to overburden pressure, which increases linearly with depth
(dotted blue lines in Figure~\ref{fig:Crack-tip-stress}), such that
the net stress at the crack tip is
\begin{equation}
\sigma_{\text{tip}}\left(y\right)=\frac{\sigma_{\text{crit}}T_{e}^{2}}{\left(T_{e}-y\right)^{2}}-\rho_{\text{ice}}gy.\label{eq:sigma_tip}
\end{equation}

The gradient as a function of depth is then given by
\begin{equation}
\frac{d}{dy}\sigma_{\text{tip}}\left(y\right)=\frac{2\sigma_{\text{crit}}T_{e}^{2}}{\left(T_{e}-y\right)^{3}}-\rho_{\text{ice}}g\label{eq:sigma_tip_gradient}
\end{equation}
such that $\sigma_{\text{tip}}\left(y\right)$ increases monotonically
with $y$ as long as %
\begin{equation}
T_{e}g<\frac{2\sigma_{\text{crit}}}{\rho_{\text{ice}}}.\label{eq:sigma_tip_increasing_with_y}
\end{equation}

Provided that (\ref{eq:sigma_tip_increasing_with_y}) is satisfied,
the tensile stresses due to bending always exceed the compressive
stresses due to overburden pressure, such that the net stresses at
the crack tip are always tensile, and increasingly so as the crack
penetrates deeper through the lithosphere (Figure~\ref{fig:Crack-tip-stress}).
As a result, once initiated in this way, crack propagation cannot
be arrested at any point within the lithosphere. Whereas the condition
described by (\ref{eq:sigma_tip_increasing_with_y}) is readily satisfied
for Enceladus due to its thin ice shell and small surface gravity,
it may not be satisfied for larger bodies or when a thicker elastic
layer is present. For example, this condition is not satisfied for
Europa unless the elastic layer thickness is at the low end of the
estimated range---less than about $\unit[1.5]{km}$ (Figure~\ref{fig:Crack-tip-stress-gradient};
\citealp{Billings2005}). For Ganymede, Callisto, and Titan, the gravity
is so strong that the compressive stresses overwhelm the bending stresses
for plausible shell thicknesses, precluding rapid crack propagation.
Hence, because this situation is permitted only for bodies with low
surface gravity, or when the ice shell is extremely thin, Enceladus
may be unique among ocean worlds for its ability to develop through-going
fractures due to bending stresses (positive or green region of Figure~\ref{fig:Crack-tip-stress-gradient}).

What happens when the crack tip reaches the ductile part of the ice?
The relatively warmer ($T>\unit[160]{K}$) lower part of the ice shell
is ductile on long timescales, and therefore does not participate
in the gradually accumulated bending stresses that support the load
at the edge of the previously broken plate. If the new fracture is
propagating sufficiently rapidly, however, the entire ice shell behaves
elastically (the Maxwell time is at least a few hours, even for the
low viscosity ice near its melting temperature). That is, as the fracture
propagates downward, stresses may build within the otherwise ductile
regions of the ice shell, helping to support the load, at least transiently.
Determining how this changes the evolution of the stress profile during
crack propagation is not trivial and will be the subject of future
work.

\section*{Supplementary Information}

\end{document}